\documentclass[twocolumn,aip,reprint,amsmath,amssymb]{revtex4-1}
\usepackage[hidelinks]{hyperref}
\usepackage{graphicx}
\usepackage{epstopdf}
\usepackage{color}
\usepackage{bm}
\usepackage{amssymb}
\usepackage{amsfonts}
\usepackage{amsmath}
\usepackage{float}
\usepackage{siunitx}
\usepackage{multirow}

\begin{document}

\title{The role of spin-orbit interaction in low thermal conductivity of Mg$_3$Bi$_2$}

\author{Nguyen Tuan Hung}
\email{nguyen.tuan.hung.e4@tohoku.ac.jp}
\affiliation{Frontier Research Institute for Interdisciplinary Sciences, Tohoku University, Sendai 980-8578, Japan}


\begin{abstract}
Three-dimensional layered Mg$_3$Bi$_2$ has emerged as thermoelectric material due to its high cooling performance at ambient temperature, which benefits from its low lattice thermal conductivity and semimetal character. However, the semimetal character of Mg$_3$Bi$_2$ is sensitive to spin-orbit coupling (SOC). Thus, the underlying origin of low lattice thermal conductivity needs to be clarified in the presence of the SOC. In this work, the first-principles calculations within the two-channel model are employed to investigate the effects of the SOC on the phonon-phonon scattering on the phonon transport of Mg$_3$Bi$_2$. Our results show that the SOC strongly reduces the lattice thermal conductivity (up to $\sim 35$ \%). This reduction originates from the influence of the SOC on the transverse acoustic modes involving interlayer shearing, leading to weak interlayer bonding and enhancement anharmonicity around 50 cm$^{-1}$. Our results clarify the mechanism of low thermal conductivity in Mg$_3$Bi$_2$ and support the design of Mg$_3$Bi$_2$-based materials for thermoelectric applications.
\end{abstract}

\date{\today}
\maketitle

The demand for green energy with net-zero gas emissions requires the development of sustainable energy-related technologies, in which thermoelectricity is one of the promising technologies that can convert heat energy into electrical energy without gas emissions. A thermoelectric (TE) device is mainly fabricated from TE material, which takes nearly one-third of the total cost of the device~\cite{leblanc2014material}. Some of the best TE materials are Bi$_2$Te$_3$, PbTe, and their related alloys, which were discovered around the 1950s and used as commercial TE materials~\cite{goldsmid1954use,heremans2008enhancement}. However, these materials are limited for wide applications due to the rare and expensive of the Te element. Therefore, during the past decade, there have been significant efforts to search for non-Te materials, such as $\alpha$-MgAgSb~\cite{zhao2014high}, Mg$_3$Bi$_2$~\cite{mao2019high,imasato2019exceptional,shi2021compromise,liu2022maximizing}, Bi$_2$Se$_3$~\cite{wang2016high}, and SnSe~\cite{zhao2014ultralow} crystals. Among them, Mg$_3$Bi$_2$ crystal is an interesting material to study the fundamental transport properties since it shows not only a high cooling performance with a large temperature difference of $\sim 91$ kelvin~\cite{mao2019high} but also topological character~\cite{chang2019realization,hung2022enhanced}.

In previous work~\cite{hung2022enhanced}, we showed that a spinless Mg$_3$Bi$_2$ could be a type-II nodal line semimetal, in which the conduction and valence bands
intersect in the form of a line (called the nodal line)~\cite{fu2019dirac}. This feature leads to van Hove singularities near the nodal line energy and enhances the TE power factor~\cite{hung2022enhanced}. However, the nodal line character of Mg$_3$Bi$_2$ is suppressed by the spin-orbit coupling (SOC), which often happens with the Bi element. By considering the SOC, Mg$_3$Bi$_2$ becomes a normal semimetal with a tiny band gap. Then, the electronic transport properties change significantly because of the missing van Hove singularities~\cite{hung2022enhanced}. On the other hand, the SOC also can affect the thermal conductivity of the materials. Tian \textit{et al.}~\cite{tian2012phonon} have reported that the phonon lifetimes (or anharmonicity) of PbSe and PbTe are larger with the SOC, which leads to twice larger thermal conductivity with the SOC than that without the SOC at room temperature. Wu \textit{et al.}~\cite{wu2019coupling} also showed that the SOC leads to enhanced lattice thermal conductivity of SnSe (up to $\sim 60$\%) compared without the SOC. On the other hand, Li \textit{et al.}~\cite{li2012thermal} showed that the SOC does not affect the lattice thermal conductivity of Mg$_2$Si and Mg$_2$Sn due to the relatively small discrepancies between their calculations and the experimental data. These studies suggest that the SOC will play an essential role in the thermal transport in Mg$_3$Bi$_2$. The previous reports~\cite{zhu2022giant,peng2018unlikely} focus only on the thermal properties of Mg$_3$Bi$_2$ without SOC. Thus, the role of the SOC in the low thermal conductivity of Mg$_3$Bi$_2$ still needs to be clarified to better understand the thermoelectric properties of Mg$_3$Bi$_2$. It is noted that we can not suppress the intrinsic spin-orbit interaction in the materials. Thus, it is difficult to observe the effect of SOC on the lattice thermal conductivity by experiment. In this situation, a theoretical calculation needs to be performed first to investigate the lattice thermal conductivity of Mg$_3$Bi$_2$ with both cases of the SOC and without SOC. 

In this Letter, we investigate the lattice thermal properties of Mg$_3$Bi$_2$ with and without SOC to clarify the influence of the SOC on phonon dispersion and lattice thermal conductivity, $\kappa_l$. By using the phonon Boltzmann transport with first-principle calculations, we found that the SOC reduces $\kappa_l$ by about 35\%, while it was reported to enhance $\kappa_l$ of PbTe, PbSe, and SnSe~\cite{tian2012phonon,wu2019coupling}. Furthermore, the first-principle calculations within the phonon-phonon interaction underestimate $\kappa_l$ compared with experimental data. Thus, we applied the two-channel model for $\kappa_l$, which accounts for the correction term by the Cahill-Watson-Pohl (CWP) formula~\cite{cahill1992lower}. 

\begin{figure}[t!]
  \centering
  \includegraphics[clip,width=7.5cm]{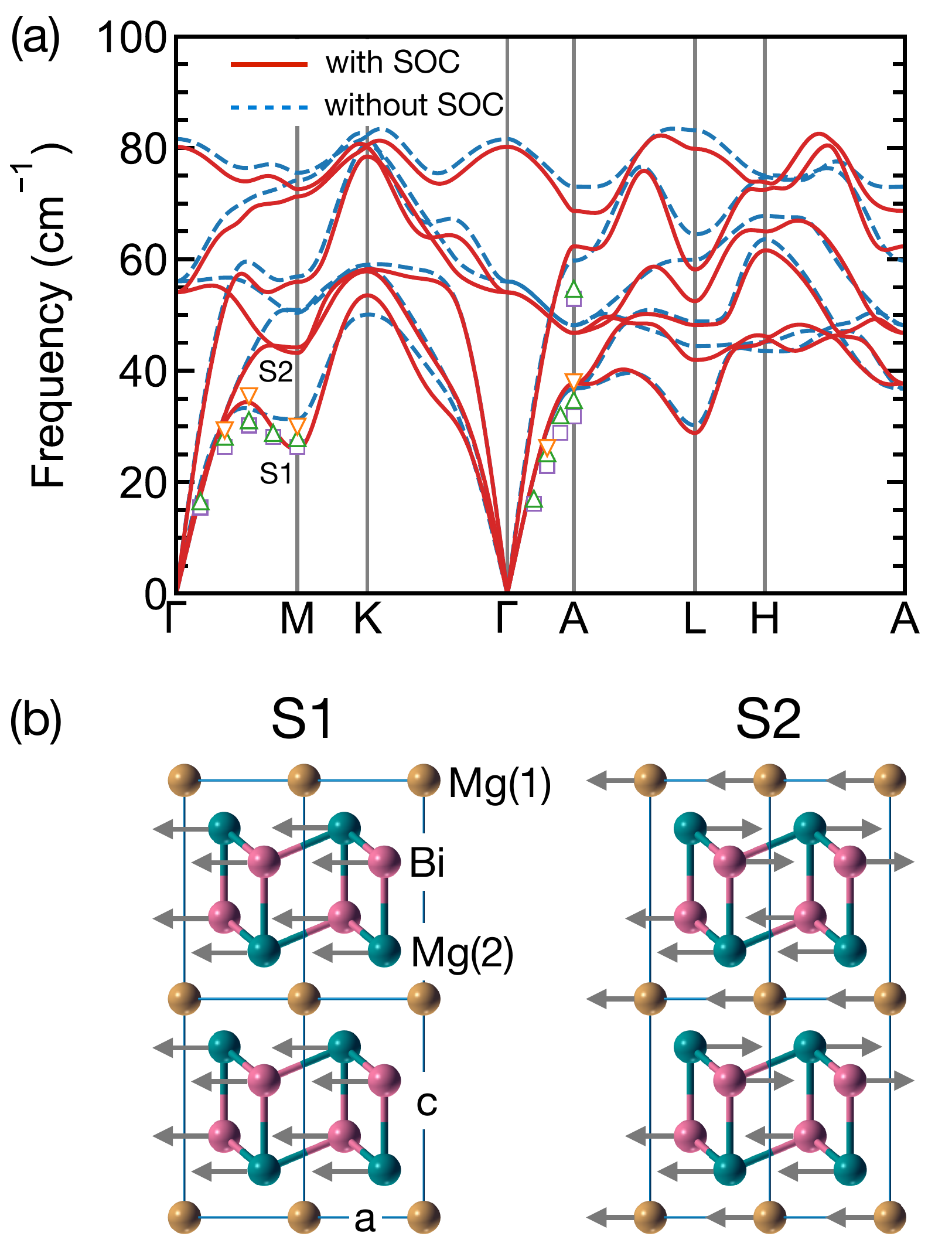}
  \caption{(a) Phonon dispersions of Mg$_3$Bi$_2$ along the high-symmetry points in the low-frequency regime with SOC (red solid line) and without SOC (blue dashed line). The symbols represent experimental data by inelastic x-ray scattering measurements~\cite{ding2021soft}, in which $\bigtriangledown$, $\triangle$, and $\square$ markers correspond to the phonon frequencies at 80, 300, and 600 K. (b) Atomic displacements of Mg$_3$Sb$_2$ correspond to the transverse acoustic phonon modes S1 $\sim 30$ cm$^{-1}$ and S2 $\sim 50$ cm$^{-1}$ at the M point, which are related to the shearing modes between the Bi-Mg(2) and Mg(1) layers.}
\label{fig:phonon}
\end{figure}

The phonon dispersion is calculated by the density-functional-perturbation theory (DFPT) with the Quantum ESPRESSO package~\cite{baroni2001phonons,giannozzi2009quantum,hung2022quantum}. The fully-relativistic and scalar-relativistic ultra-soft pseudopotentials with the Perdew-Burke-Ernzerhof (PBE) functional~\cite{perdew1996generalized} use for the calculations with the SOC and without the SOC, respectively. All positions of the atoms and lattice constants are optimized by the BFGS quasi-newton algorithm~\cite{hung2022quantum}, in which the convergence values for the forces and stress components are $0.0001$ Ry/a.u.$^3$ and $0.005$ GPa, respectively. The obtained lattice constants of Mg$_3$Bi$_2$ are $a=b=4.683$ \AA\ and $c=7.396$ \AA, which are consistent with the previous works~\cite{hung2022enhanced,zhu2022giant}. Cutoff energy of 60 Ry, $\mathbf{k}$-points mesh of $10\times 10 \times 6$, and $\mathbf{q}$-points mesh of $5\times 5 \times 3$ for all calculations are selected based on the convergence test.

In Fig.~\ref{fig:phonon}(a), we show the phonon dispersions of Mg$_3$Bi$_2$ with the SOC (solid line) and without SOC (dashed line). Only the low-frequency regime is plotted to see the difference between the solid and dashed lines easily. We noted that the high-frequency regime above the phonon band gap could contribute less than 10\% to the total thermal conductivity~\cite{dong2019new}. The phonon dispersions reproduce the inelastic x-ray scattering (IXS) spectra~\cite{ding2021soft}, in which the case of the SOC shows a better fitting with the experimental data for the soft phonon S1 around 26 cm$^{-1}$ at the M point. The main difference between the phonon frequency with the SOC and without the SOC is found at S1 ($\sim 5$ cm$^{-1}$) and S2 ($\sim 8$ cm$^{-1}$) at the M point, as shown in Fig.~\ref{fig:phonon}(a). The S1 and S2 phonon modes are the interlayer shearing modes in the Mg$_3$Bi$_2$ with $P\overline{3}m1$ space group~\cite{ding2021soft,peng2018unlikely}, as shown in Fig.~\ref{fig:phonon}(b). The Mg$_3$Bi$_2$ structures consist of alternating [Mg(2)$_2$Bi$_2$] and [Mg(1)] atom layers. We thus expect weak bonding between [Mg(2)$_2$Bi$_2$] and [Mg(1)] layers, resulting in a small shear strength. Here, we calculate the elastic modulus, including the bulk $B$, Young $E$, and shear $G$ modulus, using the Voigt-Reuss-Hill approximation~\cite{anderson1963simplified,hung2017three} with the Thermo\_pw code~\cite{dal2016elastic}. The obtained results for the cases with and without the SOC are listed in Table~\ref{table:1}, which are consistent with the experiment data ($B=38.39$ GPa, $G=13.39$ GPa, and $E=35.98$ GPa using resonant ultrasound spectroscopy~\cite{peng2018unlikely}). The shear modulus of Mg$_3$Bi$_2$ is much softer than compounds with similar structures, such as CaMg$_2$As$_2$, YbMg$_2$Sb$_2$ or BaMg$_2$P$_2$~\cite{peng2018unlikely}, resulting in soft phonon modes of S1 and S2. Another soft phonon mode related to the interlayer shearing is found around 30 cm$^{-1}$ at the $L$ point, as shown in Fig.~\ref{fig:phonon}(b). However, the SOC does not affect this phonon mode.

\begin{table}[t]
\caption{Bulk $B$, shear $G$, and Young $E$ modulus (GPa), Poisson ratio $r$, and longitudinal $v_l$ and transverse sound velocities $v_t$ (m/s) of Mg$_3$Bi$_2$.}
\centering    
{\renewcommand{\arraystretch}{1.5}    
\small
\begin{tabular}{ccccccc}
\hline\hline
Mg$_3$Bi$_2$ & $B$ & $G$ & $E$ & $r$ & $v_l$ & $v_t$ \\ 
\hline    
With SOC & 31.81 & 16.44 & 42.07 & 0.28 & 3086.99 & 1707.55 \\  
\hline  
Without SOC & 35.44 & 18.29 & 46.81 & 0.28 & 3247.72 & 1795.77 \\  
\hline\hline
\end{tabular}
}
\label{table:1}    
\end{table}	

In order to investigate the effect of the SOC on the lattice thermal conductivity $\kappa_l$ of Mg$_3$Bi$_2$, we calculate the two-channel model~\cite{mukhopadhyay2018two,du2023low,luo2020vibrational}, which is defined as follows
\begin{equation}
\label{eq:1}
\kappa_l=\kappa_{\text{ph}}+\kappa_{\text{diff}},
\end{equation}
where $\kappa_{\text{ph}}$ is the phonon channel, which is defined by~\cite{hung2019designing}
\begin{equation}
\label{eq:2}
\kappa_{\text{ph}}=\frac{1}{N_{\bm{q}}V}\sum_{\nu\bm{q}}\hbar\omega_{\nu\bm{q}}v_{\nu\bm{q}}^2\tau_{\nu\bm{q}}\frac{\partial n_{\nu\bm{q}}}{\partial T},
\end{equation}
where $N_{\bm{q}}$ is the number of $\bm{q}$ points and $V$ is the volume of the unit cell. $\omega_{\nu\bm{q}}$, $v_{\nu\bm{q}}$, and $\tau_{\nu\bm{q}}$ is the phonon frequency, the phonon group velocity, and the phonon lifetime of the phonon mode $\nu$ at $\bm{q}$ vector, respectively. $n_{\nu\bm{q}}=(e^{\hbar\omega_{\nu\bm{q}}/k_BT}-1)^{-1}$ is the Bose-Einstein distribution function. Here, $\kappa_{\text{ph}}$ and $\tau_{\nu\bm{q}}$ are calculated by solving the phonon Boltzmann transport equation, as implemented in the ShengBTE code~\cite{li2014shengbte} using $16 \times 16 \times 16$ integration meshes, based on the second-order force constants calculated the DFPT~\cite{baroni2001phonons,giannozzi2009quantum} and third-order force constants calculated with a $3\times 3 \times 3$ supercell and up to the five-nearest neighbors using thirdorder.py~\cite{li2014shengbte}. $\kappa_{\text{diff}}$ is the diffusion channel in the disordered crystal, which is described by the Cahill-Watson-Pohl (CWP) model as~\cite{cahill1992lower}
\begin{equation}
\label{eq:3}
\kappa_{\text{diff}}=\left(\frac{\pi}{6}\right)^{1/3}k_B\rho^{2/3}\sum_i v_i\left(\frac{T}{\theta_i}\right)^{2}\int\limits_0^{\theta_i/T}\frac{x^3e^x}{(e^x-1)^2}\mathrm{d}x,
\end{equation}
where $\rho$ is the number density of atoms, $x$ is the dimensionless integration variable, $\theta_i=v_i(\hbar/k_B)(6\pi^2\rho)^{1/3}$ is the Debye temperature, where $v_i$ is the sound speed of acoustic branch $i$, including one longitudinal sound velocity $v_L = \sqrt{(B+4G/3)/\rho}$ and two transverse sound velocity $v_T = \sqrt{G/\rho}$. The calculated $v_L$, $v_T$, $B$ and $G$ are given in Table~\ref{table:1}.

\begin{figure}[t!]
  \centering
  \includegraphics[clip,width=7.7cm]{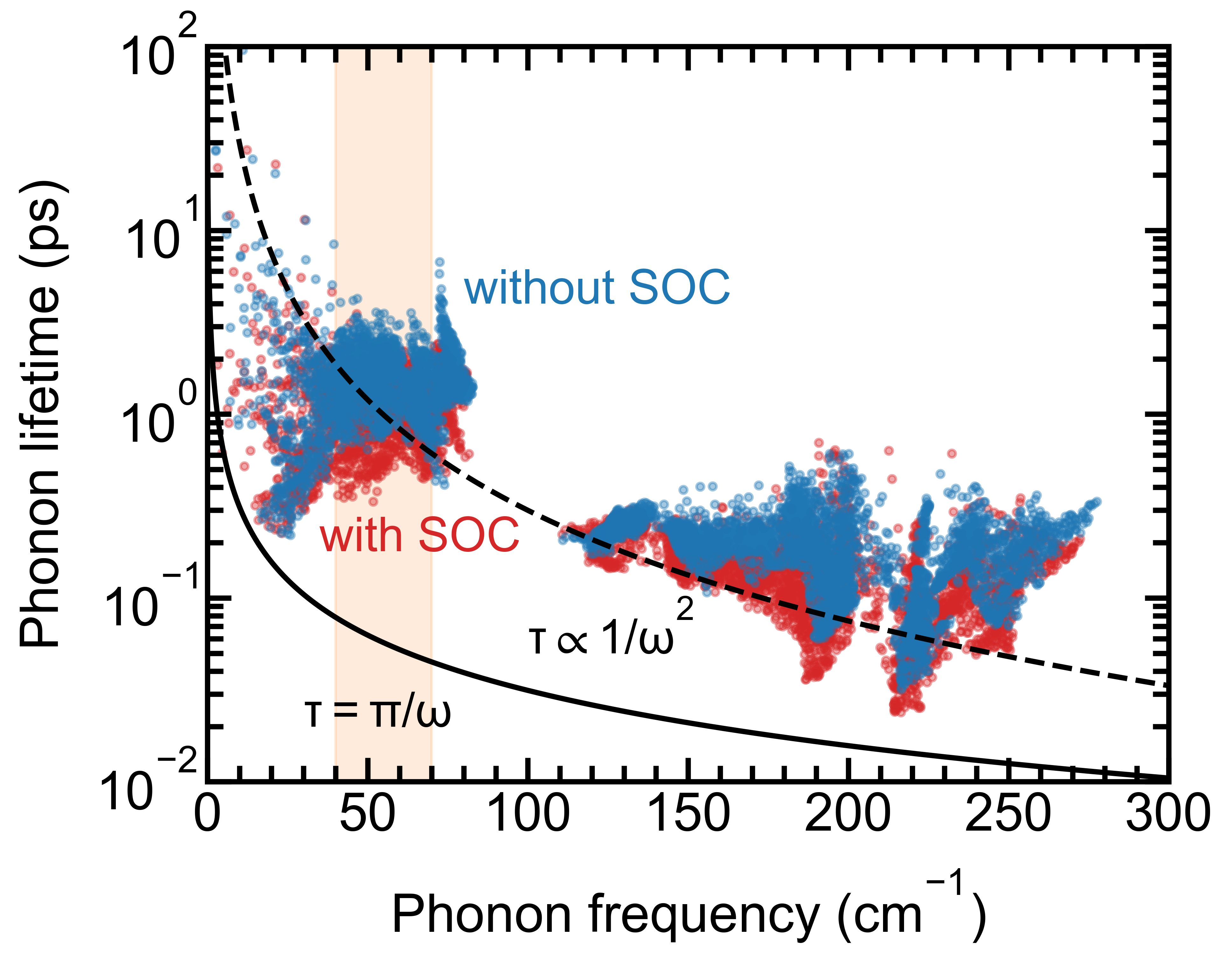}
  \caption{Phonon lifetime $\tau$ of Mg$_3$Bi$_2$ at room temperature $T=300$ K is plotted as a function of phonon frequency $\omega$ with SOC (red dots) and without SOC (blue dots). The black solid and dashed cures give the minimum lifetime $\tau=\pi/\omega$ from the CWP formula and $\tau\propto 1/\omega^2$ from the Umklapp scattering, respectively.}
\label{fig:lifetimes}
\end{figure}

In Fig.~\ref{fig:lifetimes}, we show the phonon lifetime at $T=300$ K of Mg$_3$Bi$_2$ with and without the SOC. The effect of the SOC on the phonon lifetime is considered only for $\kappa_{{\text{ph}}}$, in which the phonons frequencies below $\omega < 80$ cm$^{-1}$ mainly dominate on $\kappa_{{\text{ph}}}$ due to the strong frequency dependence of the Umklapp scattering ($\tau \propto 1/\omega^2$)~\cite{hung2019designing}. For $80 < \omega < 110$ cm$^{-1}$, there is no phonon lifetime because of the phonon band-gap region. For $\omega > 110$ cm$^{-1}$, the average value of phonon lifetime is about 0.1 ps, which is one order of magnitude smaller than that for $ \omega <80$ cm$^{-1}$ ($\sim 1$ ps). Thus, the contribution of the SOC to $\kappa_{{\text{ph}}}$ becomes an important factor when $\omega < 80$ cm$^{-1}$.
In particular, the SOC leads to a reduction in the phonon lifetime around $50$ cm$^{-1}$, which corresponds to the interlayer shearing mode S2, as shown in Fig.~\ref{fig:phonon}(b). We note that the high anharmonicity (i.e., low phonon lifetime) is also found at $\omega \time 20-30$ cm$^{-1}$. This is the contribution of both the shearing mode S1 at the M point and another shearing mode ($\sim$ $30$ cm$^{-1}$) at the L point (see Fig.~\ref{fig:phonon}(a)). Since the SOC does not affect the shearing mode at the L point, the phonon lifetime with $\omega < 30$ cm$^{-1}$ does not change significantly with the presence of the SOC.
Besides that, the SOC also does not affect the phonon lifetime in the CWP model since the CPW model assumes that the phonon lifetime is half the period of oscillation~\cite{mukhopadhyay2018two} (i.e., $\tau=\pi/\omega$).

\begin{figure}[t!]
  \centering
  \includegraphics[clip,width=7.5cm]{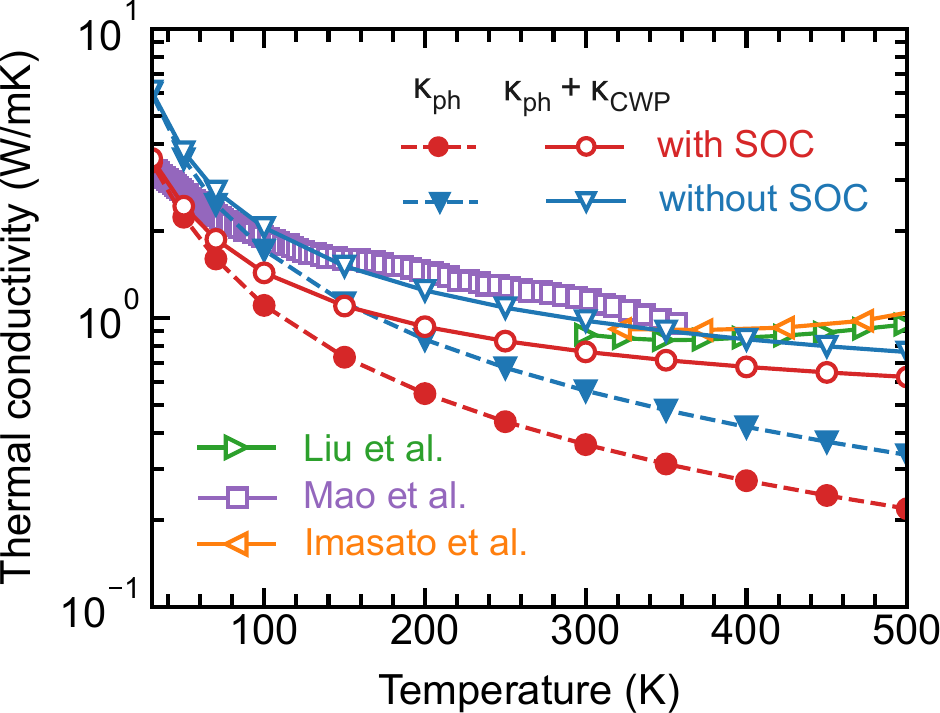}
  \caption{Thermal conductivity of Mg$_3$Bi$_2$ with and without SOC is plotted as a function of the temperature $T$. Symbols $\blacktriangleright$, $\blacksquare$, and $\blacktriangleleft$ represent the experimental data from Liu \textit{et al.}~\cite{liu2022maximizing}, Mao \textit{et al.}~\cite{mao2019high} and Imasato \textit{et al.}~\cite{imasato2019exceptional}, respectivity.}
\label{fig:thermal}
\end{figure}

In Fig.~\ref{fig:thermal}, we show the thermal conductivity $\kappa_l$ of Mg$_3$Bi$_2$ as a function of the temperature $T$. The obtained $\kappa_{\text{ph}}$ is almost isotropic in Mg$_3$Bi$_2$. At 300 K, $\kappa_{\text{ph}}^{xx} = \kappa_{\text{ph}}^{yy} = 0.37; \kappa_{\text{ph}}^{zz} = 0.36$ W/mK for the case with SOC and $\kappa_{\text{ph}}^{xx} = \kappa_{\text{ph}}^{yy} = 0.53; \kappa_{\text{ph}}^{zz} = 0.59$ W/mK for the case without SOC. Thus, the average value of $\kappa_{\text{ph}}=(\kappa_{\text{ph}}^{xx}+\kappa_{\text{ph}}^{yy}+\kappa_{\text{ph}}^{zz})/3$ is plotted in Fig.~\ref{fig:thermal}. We note that the Mg$_3$Bi$_2$ has a 3D layered structure, and it shows a two-dimensional (2D) electron character, i.e., electrons mostly move in [Mg(2)$_2$Bi$_2$] layer~\cite{hung2022enhanced}. However, using quantitative analysis of chemical bonding, Zhang \textit{et al.}~\cite{zhang2018chemical} showed that the interlayer and intralayer bonds of Mg$_3$Bi$_2$ are largely ionic with partial covalent nature. Thus, Mg$_3$Bi$_2$ exhibits a nearly isotropic three-dimensional (3D) bonding network, leading to mostly isotropic $\kappa_{\text{ph}}$. Interestingly, such 2D-electron and 3D-phonon transports of Mg$_3$Bi$_2$ are opposite to 3D-electron and 2D-phonon transports of SnSe~\cite{chang20183d}. We can see that $\kappa_{\text{ph}}$ is reduced by about 35\% 
enhanced anharmonicity of the S2 shearing mode. Since the SOC affects the shearing modulus, it also reduces the longitudinal and transverse sound velocities (see Table~\ref{table:1}). Thus, $\kappa_{\text{diff}}$ is also reduced by the SOC ($\kappa_{\text{diff}}=0.40$ and 0.42 W/Km with SOC and with SOC, respectively). As shown in Fig.~\ref{fig:thermal}, $\kappa_{\text{ph}}$ much lower than the experimental observation~\cite{mao2019high,liu2022maximizing,imasato2019exceptional} when $T>50$ K due to neglect of the temperature-dependent anharmonic renormalization of the phonon frequencies~\cite{xia2020particlelike}. By using $\kappa_{\text{diff}}$ to correct this term, $\kappa_{\text{ph}}+\kappa_{\text{diff}}$ can reproduce the experimental observation~\cite{mao2019high,liu2022maximizing,imasato2019exceptional}. 

In conclusion, we have employed a two-channel model to study the effects of the SOC on the phonon dispersion, phonon anharmonicity, and thermal conductivity of Mg$_3$Bi$_2$. Our calculations reproduce well the experimental data. The SOC not only enhances the anharmonicity of the interlayer shearing mode but also reduces the longitudinal and transverse sound velocities. Therefore, the SOC can have a considerable impact on thermal transport properties. Our calculations suggest a potential way for manipulating phonon transport by tuning the SOC. 

\section*{Acknowledgments}
N.T.H. acknowledges financial support from the Frontier Research Institute for Interdisciplinary Sciences, Tohoku University.

\section*{References}
\bibliographystyle{aip}
%

\end{document}